\documentclass[11pt]{article}  

\newcommand{\CHAP}[2][]{#1} %

\usepackage[letterpaper, left=1in, right=1in, top=1in, bottom=1in]{geometry}
\usepackage[utf8]{inputenc} %
\usepackage[T1]{fontenc}    %
\usepackage{url}            %
\usepackage{microtype}      %
\usepackage{xcolor}         %
\usepackage{csquotes}       %
\MakeOuterQuote{"}

\usepackage{graphicx}
\usepackage{graphbox}
\usepackage{caption}
\usepackage{subcaption}
\usepackage{float}
\usepackage{booktabs}

\usepackage{amsfonts}
\usepackage{amsmath}
\usepackage{amssymb}
\usepackage{bbm}
\usepackage{bm}

\usepackage[hidelinks]{hyperref}  %

\usepackage{cleveref} %
\crefname{figure}{Figure}{Figures}

\usepackage[style=apa]{biblatex}
\renewcommand{\v}{\bm}

\renewcommand{\th}{\textsuperscript{th}}

\newcommand{\softmax}{\mathrm{softmax}}
\newcommand{\sgn}{\mathrm{sign}}

\newcommand{\ie}{\textit{i.e.}}
\newcommand{\eg}{\textit{e.g.}}

\title{Computational models of learning and synaptic plasticity}
\author{Danil Tyulmankov\textsuperscript{1,2,}\thanks{Correspondence: tyulmank@usc.edu}\\
\footnotesize\textsuperscript{1}Center for Theoretical Neuroscience, Columbia University, New York, NY, USA\\
\footnotesize\textsuperscript{2}Viterbi School of Engineering, University of Southern California, Los Angeles, CA, USA}
\date{}

\begin{document}
\maketitle


\section*{Learning objectives}
\begin{itemize}
    \item Summarize the key models of synaptic plasticity and types of signals that can be used to update synaptic weights.
    \item Provide an outline for organizing several major classes of learning and memory paradigms.
    \item Link learning outcomes at a behavioral level to synaptic plasticity mechanisms that can implement them. 
\end{itemize}

\section*{Abstract}
Many mathematical models of synaptic plasticity have been proposed to explain the diversity of plasticity phenomena observed in biological organisms. These models range from simple interpretations of Hebb's postulate, which suggests that correlated neural activity leads to increases in synaptic strength, to more complex rules that allow bidirectional synaptic updates, ensure stability, or incorporate additional signals like reward or error. At the same time, a range of learning paradigms can be observed behaviorally, from Pavlovian conditioning to motor learning and memory recall. Although it is difficult to directly link synaptic updates to learning outcomes experimentally, computational models provide a valuable tool for building evidence of this connection. In this chapter, we discuss several fundamental learning paradigms, along with the synaptic plasticity rules that might be used to implement them.    

\newpage
\renewcommand*\contentsname{Outline}
\tableofcontents
\newpage

\section{Introduction}
Synaptic plasticity is regarded as the physical basis for learning and memory in animals \parencite{kandel2014molecular}. It is broadly categorized into two main types: short-term and long-term synaptic plasticity \parencite{citri_synaptic_2008}. Short-term plasticity, as the name suggests, occurs on timescales from milliseconds to minutes \parencite{zucker_short-term_2002}. It can result from transient changes in presynaptic calcium levels, which alter neurotransmitter release probability, or from postsynaptic receptor desensitization due to prolonged neurotransmitter exposure. Short-term plasticity may provide a circuit with additional computational capabilities which can extend or enhance those already implemented by neuronal dynamics. For instance, short-term plasticity can function as a filter for information transmission \parencite{abbott_synaptic_2004} or contribute to working memory storage \parencite{mongillo_synaptic_2008, masse_circuit_2019}.

Long-term plasticity, on the other hand, causes changes that can last for hours, days, or years. As with short-term plasticity, long-term plasticity is bidirectional, and can cause either a strengthening (long-term potentiation, LTP) or weakening (long-term depression, LTD) of a synapse. A plethora of biochemical processes regulate the synaptic strength \parencite{malenka2004ltp}, mainly through the upregulation or downregulation of postsynaptic AMPA receptors. This regulation is dependent on the simultaneous activity of pre- and postsynaptic neurons and is mediated by NMDA receptors, which detect coincident activity \parencite{luscher2012nmda}. Long-term plasticity is widely regarded as the primary physical mechanism underlying learning and memory in animals \parencite{langille_synaptic_2018, takeuchi_synaptic_2014, martin_synaptic_2000, martin_new_2002}. Consequently, modeling it will be the main focus of this chapter.

The strength of a synapse can be estimated through several measures, such as the size of the dendritic spine, the number of synaptic receptors, the slope of the postsynaptic potential in response to a presynaptic spike, or the amplitude of a postsynaptic current \parencite{glasgow2019approaches}. In computational models, however, the strength of a synapse is commonly denoted by a single scalar $w_{ij}$, known as the synaptic weight from a presynaptic neuron $j$ to a postsynaptic neuron $i$. Our goal is then to establish the rules that govern the temporal evolution of the synaptic weights -- driven by neural activity evoked by memory or exteroceptive or interoceptive stimuli -- and its consequent impact on an animal's behavior. 

With modern optical and electrophysiological techniques, we can simultaneously measure the activity of thousands of neurons \textit{in vivo}. However, it remains experimentally challenging to measure the temporal evolution of even a single synapse. Attempting to do so simultaneously for the multitude of synaptic weights in a neural circuit together with the neurons that drive their plasticity in a live behaving animal has been particularly out of reach \parencite{humeau_next_2019}, with only very recent developments making progress on this front \parencite{gonzalez2023synaptic}. Therefore, although computational modeling is an invaluable tool for neuroscience overall, it is particularly indispensable for studying the synaptic mechanisms of learning and memory. 

Computational models allow us to explore plasticity rules in a fully controlled and observable way, free from the limitations of recording technology or behavioral experiments. Given access to both the neurons and synapses in a model circuit, we can, for example, investigate how the synaptic dynamics, defined by the plasticity rule, impact the neural network dynamics, which are defined in turn by the synaptic connectivity \parencite{clark2024theory}. Furthermore, while synaptic plasticity is an undisputed biological phenomenon, its connection to learning is less certain. Through computational modeling, it is possible to demonstrate this link. In this case, we can search for the plasticity rule that supports a particular type of behavioral adaptation (\ie, learning). Potential rules can be discovered, for example, optimizing a plasticity rule to solve a particular learning or memory task \parencite{confavreux_meta-learning_2020, tyulmankov_meta-learning_2022}.  Alternatively, they can be inferred from neural activity \parencite{ramesh_indistinguishable_2023, lim_inferring_2015}, behavior \parencite{rajagopalan_reward_2023, ashwood_inferring_2020}, or both \parencite{mehta_model_2024}. 

Conversely, given a plasticity rule, we can examine the behaviors that arise in a model, or taking it even further, behaviors that could theoretically arise, and under which conditions. We can investigate the range of functions (\ie, behaviors) that a particular network architecture, combined with a plasticity rule, can learn, as well as the information needed for learning -- whether local neural activity is sufficient, or perhaps error signals, reward information, or other factors are required. We can calculate the memory storage capacity enabled by a particular plasticity rule, and, assuming biological rules maximize capacity, make inferences about mechanisms that are likely to exist in biological systems. This chapter focuses on such inquiries. We will provide an overview of classical models of synaptic plasticity and their application to various learning and memory paradigms,\footnote{Although, in several cases, the plasticity models and/or learning paradigms were established outside of the field of neuroscience -- usually in machine learning or artificial intelligence -- and only retrospectively applied to modeling the brain and behavior.} including recall and recognition memory, as well as supervised, unsupervised, and reinforcement learning. We will furthermore discuss more complex plasticity rules, the role of network architecture, as well as learning mechanisms that work alongside, or even in lieu of, synaptic plasticity.

\section{From abstract to biological models}

\subsection{Firing rates}

Behavioral changes in response to learning reflect changes in neural activity, which are implemented through changes in the synapses which interconnect them. Synaptic changes are, in turn, driven by neural activity, so to model plasticity we must first consider modeling the neuron itself. Simplifying the vast complexity of a neuron's spatial morphology \parencite{sjostrom_dendritic_2008} and even ignoring any temporal dynamics \parencite[cf., \eg,][]{wilson1972excitatory}, an abstract model neuron known as the McCulloch-Pitts neuron \parencite{mcculloch_logical_1943}, or its successor, the perceptron \parencite{rosenblatt_perceptron_1958}, considers only its instantaneous firing rate -- a real-valued scalar -- as a function of its inputs. 

Slightly generalizing from the original formulations, the firing rate of a postsynaptic neuron $y_i$ is computed through a weighted sum of the firing rates of its presynaptic neighbors $x_1, \ldots, x_N$, with the weights $w_{ij}$ representing synapses:
\begin{equation}
    y_i = \phi \left( \sum_{j=1}^N w_{ij} x_j \right)
\end{equation}
The function $\phi(\cdot)$ -- referred to as the neuronal nonlinearity, transfer function, or, in the parlance of modern artificial neural network models, the activation function -- may be thought of as an abstracted version of the biophysical process that converts the cell's input current $\sum_j w_{ij} x_j$ (in units of Amperes) to a firing rate (in units of spikes per second). 

At this level of abstraction, the rates of neural firing determine the synaptic update. As originally postulated by \textcite{hebb_organization_1949}, "When an axon of cell \textit{A} is near enough to excite a cell \textit{B} and repeatedly or persistently takes part in firing it, $[\ldots]$ \textit{A}’s efficiency, as one of the cells firing \textit{B}, is increased." Put more succinctly, "Cells that fire together wire together" \parencite{shatz_1992_developing}. A straightforward interpretation of this theory suggests that the change in synaptic weight $\Delta w_{ij}$ depends on correlations between the presynaptic (cell \textit{A}) and postsynaptic (cell \textit{B}) neuron's firing rates \parencite{gerstner_hebbian_2016}. The simplest implementation of this corresponds to a multiplicative relationship:
\begin{equation}\label{eq:hebb_mult}
\Delta w_{ij} = \eta x_j y_i 
\end{equation}
where $\eta>0$ is a learning rate that controls the magnitude of the update. A more general interpretation
suggests that the change in synaptic weight $\Delta w_{ij}^{(t)}$ at any given time $t$ depends on the entire history of the pre- and postsynaptic neurons' firing rates through an arbitrary function $f(\cdot)$:
\begin{equation}\label{eq:hebb_general}
\Delta w_{ij}^{(t)} = f\left(x_j^{(0)}, \dots, x_j^{(t)}, y_i^{(0)}, \dots, y_i^{(t)} \right)
\end{equation}
This includes the most well-known plasticity rule of \autoref{eq:hebb_mult}, often referred to as "the" Hebbian rule, as well as other forms of Hebbian rules involving higher-order interactions between the pre- and postsynaptic activity; more complex rules such as the covariance \parencite{sejnowski_storing_1977, willshaw_optimal_1990} or BCM \parencite{bienenstock_theory_1982} rules; "non-Hebbian" plasticity rules that are driven by the presynaptic neurons alone or postsynaptic neurons alone; rules with homeostatic components such as weight decay; and even "anti-Hebbian" rules with a negative learning rate $\eta<0$. This class of models of synaptic plasticity is the focus of this chapter, but we first briefly discuss models relying on a lower level of abstraction. 

\subsection{Spikes}
In reality, rather than smoothly varying firing rates, neurons communicate through discrete action potentials or "spikes" \parencite{gerstner_neuronal_2014}. Peeling away the firing-rate layer of abstraction \parencite{abbott_model_1990}, we can consider plasticity rules that rely directly on the spiking activity \parencite{taherkhani_review_2020}. A major class of such rules, known as spike-timing dependent plasticity (STDP), depends on the precise timing of the spikes to determine the changes in synaptic strength \parencite{caporale2008spike, morrison_phenomenological_2008}. These rules are also part of the Hebbian family of plasticity rules, as they are local to the synapse and depend on correlations between the pre- and postsynaptic neurons, or more generally can be described by an equation analogous to \autoref{eq:hebb_general}. In fact, in some cases, it is possible to directly reduce a spike-timing based rule to a rate-based one\footnote{And vice versa, to include timing effects in rate-based models \parencite[Ch. 8]{dayan_theoretical_2005}.} via averaging \parencite{kempter1999hebbian}. 

The simplest instantiation of STDP is based on pair-wise interactions of spikes \parencite{gerstner1996neuronal, song2000competitive}. This rule suggests that a synapse is potentiated if the presynaptic neuron fires a spike before the postsynaptic one (following Hebb's postulate that cell \textit{A} "takes part in firing" cell \textit{B}) and depressed if vice versa (Hebb did not include a prescription for synaptic depression, but this idea has indeed found experimental support). Although \textit{a priori} the learning window can be any shape, it has been experimentally demonstrated in some cases \parencite{bi1998synaptic, zhang1998critical}, that the change in weight is an exponential function of the time difference:
\begin{equation}\label{eq:stdp}
    \Delta w_{ij} = 
        \begin{cases}
            A_p \exp{\left(\frac{t_j - t_i}{\tau_p}\right)} &\text{ if } t_i > t_j\\
            -A_d \exp{\left(\frac{t_j - t_i}{\tau_d}\right)} &\text{ if } t_i < t_j \\
        \end{cases}
\end{equation}
where $A_p, \tau_p$ and $A_d, \tau_d$ are positive constants determining the rates of potentiation or depression, and $t_i, t_j$ are the times of the pre- and postsynaptic spikes, respectively.  

A wide variety of more complex spike-timing rules have been modeled and observed experimentally \parencite{feldman2012spike}, including anti-Hebbian STDP, similar to \autoref{eq:stdp} with the sign flipped; rules that include homeostatic effects; rules that induce only depression or only potentiation, regardless of whether the pre- or postsynaptic neuron spiked first; rules that take into account nonlinear summation effects of multiple spikes, such as triplet-based STDP \parencite{pfister_triplets_2006, gjorgjieva2011triplet} or burst-dependent plasticity \parencite{payeur2021burst}; or rules which take into consideration not only the spike times but also the membrane voltage \parencite{clopath2010connectivity, clopath_voltage_2010, brader2007learning}. Overall, STDP rules are more difficult not only to analyze mathematically, but also to use in large-scale network models \parencite{abbott_building_2016}; they do, however, offer the advantage of being easier to map onto biological neurons. Nevertheless, they still do not describe the detailed biophysical processes which lead to synaptic changes, which we consider next. 

\subsection{Molecules}
The models of plasticity discussed so far are known as phenomenological models -- they simply describe the relationship between neural activity and the synaptic update. We can also consider biophysical models that assign explicit biological substrates for inducing and implementing the plasticity update \parencite{graupner_mechanisms_2010}. One molecule crucial for the induction of plasticity is calcium; according to the so-called calcium hypothesis, strong postsynaptic calcium transients drive potentiation and smaller ones drive depression \parencite{lisman1989mechanism}. We can therefore consider the change in synaptic weight to be a function of calcium concentration $c_{ij}^{(t)}$ at the synapse:
\begin{equation}
    \Delta w_{ij}^{(t)} = f(c_{ij}^{(t)})
\end{equation}
The calcium dynamics $\Delta c_{ij}^{(t)}$ themselves can then be modeled "phenomenologically" -- as either linear \parencite{graupner_calcium-based_2012} or nonlinear \parencite{graupner_natural_2016} summations of exponentially decaying transients driven by pre- and postsynaptic spikes -- or "biophysically" by explicitly modeling, for example, the dynamics of NMDA receptors \parencite{shouval_unified_2002} and/or voltage-gated calcium channels \parencite{karmarkar_model_2002} which determine the calcium influx into a cell, or even considering calcium stores in the endoplasmic reticulum \parencite{mahajan2019intracellular}. 

We can also directly assign biophysical quantities to the mechanisms by which calcium concentrations drive changes in synaptic weights. For example, \textcite{abarbanel2003biophysical} suggest the dynamics of AMPA receptor phosphorylation and dephosphorylation -- themselves governed by a "phenomenological" function of calcium levels -- directly control the value of the synaptic weight. One level of biophysical realism deeper, \textcite{castellani2005model} propose a more detailed model of the enzymatic pathway that leads to this AMPA receptor phosphorylation/dephosphorylation. Other models consider the autophosphorylation of CaMKII as the substrate determining the synaptic weight \parencite{lisman1988feasibility, graupner_stdp_2007}.%

Finally, as hinted above, we note that despite their distinction as "biophysical" models, these models must nevertheless contend with some degree of abstraction. Indeed, this is true of every model, by definition; it is a matter of choosing the level of abstraction that is most useful in any particular case. Therefore, rather than contrasting biophysical and phenomenological models, it is more instructive to consider a spectrum of models from the most biophysically detailed to the most high-level, each of which is best suited for describing a different aspect of plasticity. The line denoting the level of abstraction might be drawn at the level of receptor dynamics, calcium transients, or individual spikes. Or, it might be useful to do away with the details of cellular and sub-cellular machinery and simply consider a synapse directly driven by abstract neuronal firing rates. In this chapter, we primarily work with the latter level as it is most mathematically tractable and offers practical utility through ease of designing networks that perform specific tasks. Nevertheless, many of the learning and memory paradigms discussed here have analogues implemented at a finer level of granularity, most commonly in networks governed by variations of STDP. 

\section{Memory paradigms}\label{sec:memory}

\begin{figure}
    \begin{subfigure}{0.49\textwidth}
        \centering
        \includegraphics[height=2in]{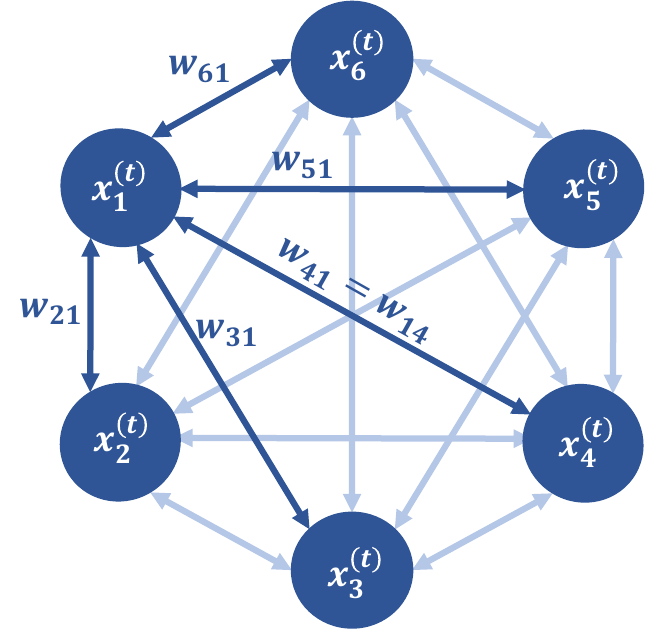}
        \caption{}    
        \label{fig:hopfield}
    \end{subfigure}
    \hfill
    \begin{subfigure}{0.49\textwidth}
        \centering
        \raisebox{0.2\height}{\includegraphics[height=1.2in]{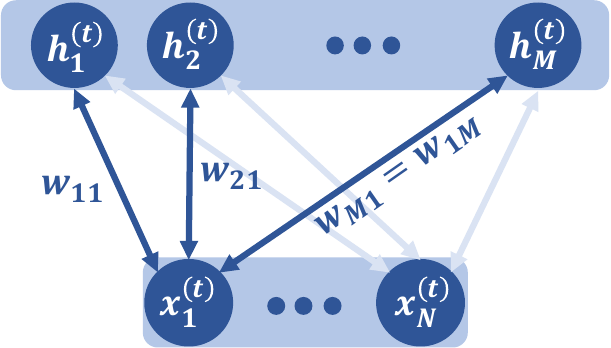}}
        \caption{}
        \label{fig:mhn}
    \end{subfigure}    
    \caption{(a) Classical associative memory, or Hopfield network with $N=6$ neurons, recurrently connected through symmetric synaptic weights $w_{ij}=w_{ji}$. (b) Dense Associative Memory, or modern Hopfield network.  A visible layer $\v{x}$ projects to a hidden layer $\v{h}$, which projects back to the visible layer through symmetric synaptic weights.}
    \label{fig:assoc_mem}
\end{figure}

\subsection{Recall}\label{sec:recall}
\subsubsection{Associative memory}
The colloquial use of “memory” commonly refers to declarative memory (also called explicit memory) -- the storage of facts (semantic memory) or experiences (episodic memory) -- which requires intentional conscious recall. One of the most influential models of recall is the associative memory network (\autoref{fig:hopfield}), also known as the Hopfield network\footnote{Similar models had been studied previously, \eg, by \textcite{little_1974_existence, little_1975_statistical, little_1978_analytic}.} \parencite{hopfield_neural_1982}. The model's objective is to store a set of items $\{\xi^1, ..., \xi^P\}$ with $\xi^p_i = \pm 1$, such that when a perturbed version $\widetilde{\xi}^p$ of one of the items is presented, the network retrieves the stored item that is most similar to it. For example, given a series of images, as well as a prompt where one of the images is partially obscured, the network would be able to reconstruct the full image. More abstractly, given a series of lived experiences, this may correspond to a verbal prompt to recall a piece of semantic or autobiographical information. 

This model is a recurrent neural network where the neural firing rate at the current timestep is defined through a weighted sum of the firing rates at the previous timestep:
\begin{equation}\label{eq:hopfield_dynamics}
    x_i^{(t+1)} = \sgn \left( \sum_{j=1}^N w_{ij} x_j^{(t)} \right)
\end{equation}
where the $\sgn(\cdot)$ function returns $+1$ if its input is positive, representing an active neuron, and $-1$ if it is negative, representing a silent neuron. To read out a memory, the network activity is initialized at time $t=0$ with a query: $\v{x}^{(0)}=\widetilde{\xi}^p$. This may come from a sensory input or from another brain area projecting to it, corresponding to an incomplete memory cue. The neurons evolve according to the dynamics in \autoref{eq:hopfield_dynamics} until they reach a "fixed point" $\v{x}^{(t)} = \v{x}^{(t+1)} = \v{x}^*$ where the update step does not change any of the neurons' firing rates. If the memory is successfully retrieved, the neural activity at the fixed point will correspond to one of the stored memories $\v{x}^* = \xi^p$. 

If the query is indeed a a firing rate pattern corresponding to an incomplete memory, these dynamics correspond to a process of "pattern completion". For this reason, the autoassociative network model is often thought of as the hippocampal area CA3, which shows long-term potentiation as a form of storing memories, has recurrent excitatory connections similar to the associative model architecture, and has been shown to perform similar pattern completion from stimulation \textit{in vitro}
\parencite{rolls_mechanisms_2013}.%

To store the given set of memories, in the classical prescription, the synaptic weights are given by:
\begin{equation}\label{eq:hopfield_weights}
    w_{ij}=\frac{1}{N} \sum_{p=1}^P \xi_i^p \xi_j^p 
\end{equation}
From a biological perspective, this can be considered to be an iterative weight update, corresponding to a Hebbian learning rule (\autoref{eq:hebb_mult}) where the neural activity $\v{x}$ is set by one of the input patterns $\xi^p$, and a learning rate of $\frac{1}{N}$ (although this coefficient is only for mathematical convenience and does not affect the network's performance):
\begin{equation}\label{eq:hopfield_update}
\Delta w_{ij}=\frac{1}{N} \xi_i^p \xi_j^p     
\end{equation}
Note, however, that this plasticity rule implies that the synaptic weights are symmetric, $w_{ij} = w_{ji}$. This is a major limitation of the Hopfield network as a model of the brain -- connections among biological neurons are rarely bidirectional in this way -- but this simplifying assumption enables analytic tractability in studying its dynamics and guarantees that the network converges to a fixed point. Non-symmetric weight matrices, on the other hand, can lead to arbitrary dynamics such as oscillations or chaos. 

\subsubsection{Hopfield network capacity}\label{sec:hopfield_capacity}
An associate memory network comprised of $N$ neurons stores a set of $P$ items for recall. A natural question arises: what is the upper limit on the number of items it can store, as a function of network size? In many cases, the answer can be calculated analytically, although the precise value depends on a number of factors such as the statistics of the dataset itself, %
the specifics of the network dynamics, %
the performance metric, %
and, importantly, the learning rule \parencite{amit_modeling_1989, hertz_introduction_1991}. 

For example, we can consider how much error is allowed during retrieval. Allowing even a small amount of error significantly improves capacity. In this case, using techniques from statistical mechanics, \textcite{amit_1985_storing} famously proved that the capacity is (assuming non-sparse and uncorrelated patterns, and that $N$ is large):
\begin{equation}
    P_{max} = 0.138N
\end{equation}
where the average fraction of incorrect elements of a retrieved pattern is less than 1.5\% for $P<P_{max}$.\footnote{This estimate was subsequently improved slightly, to $P_{max}=0.144N$ with <0.9\% errors \parencite{crisanti1986saturation} by including the effects of replica symmetry breaking.} On the other hand, fewer patterns can be stored if the retrieval is to be perfect, reducing the capacity by a factor proportional to $\frac{1}{\log(N)}$ \parencite{mceliece_capacity_1987}.

Correlations among the patterns can, in principle, also allow more patterns to be stored. Intuitively, correlations imply less unique information per pattern, so although the same amount of information is stored in the network, the number of patterns is higher. Similarly, if the patterns are biased or sparse (and therefore, by definition, correlated), each pattern can theoretically be represented with a smaller number of bits, so more patterns can fit into the network's storage. Exploiting these correlations, however, may be difficult -- a Hopfield network using the Hebbian learning rule (\autoref{eq:hopfield_update}) assumes uncorrelated patterns. If this assumption is broken, memory performance decreases dramatically, storing at most a constant number of patterns (controlled by the sparsity level), regardless of the number of neurons \parencite{amit1987information}. Remedying this to take advantage of sparsity requires modification of not only the plasticity rule but also the retrieval dynamics \parencite{amit1987information, tsodyks1988enhanced, treves1991determines}.

Critically, the capacity depends on the specific plasticity rule. Although not all are biologically plausible -- either non-local (requiring knowledge of neuronal activity beyond the pre- and postsynaptic neurons), or non-incremental (requiring knowledge of the entire dataset before setting the synaptic weights) -- there exist alternatives which enable a greater capacity. For example, the pseudo-inverse rule \parencite{personnaz_1985_information, kanter1987associative},
\begin{equation}\label{eq:pseudo_inverse_rule}
    w_{ij}=\frac{1}{N} \sum_{p=1}^P \sum_{q=1}^P \xi_i^p \v{C}^{-1}_{pq} \xi_j^q 
\end{equation}
where $\v{C}$ is the correlation matrix of the dataset, increases the associative memory network's capacity to $P_{max} = N$. This learning rule orthogonalizes the patterns and eliminates cross-talk among them during retrieval, allowing not only for a larger capacity, but also enabling storage of correlated patterns\footnote{The patterns must also be linearly independent, which is required for the inverse of $\v{C}$ to exist. The upper bound is therefore simply due to linear algebra; it is the maximum number of $N$-dimensional patterns that can be mutually orthogonal.}, as well as error-free retrieval.

In the most general case, allowing for an arbitrary (non-symmetric) synaptic weight matrix, capacity can be as high as $P_{max} = 2N$ for uncorrelated patterns, or even larger if correlations exist (but still proportional to the number of neurons, \ie, $P_{max} = \alpha N$) \parencite{gardner_space_1988}. Another consideration, however, is the size of the "basin of attraction" of a memory -- the set of initial conditions (queries $\widetilde{\xi}^p$) which converge to the memory. Larger basins of attraction imply smaller values of $\alpha$, which only approaches $2$ (for uncorrelated patterns) as the size of the basin of attraction goes to zero \parencite{gardner_space_1988}. This capacity can be achieved using a variation of the perceptron algorithm \parencite{gardner_space_1988}, but at the expense of requiring iterative error-based updates rather than a one-shot rule as in \autoref{eq:hopfield_update} or \autoref{eq:pseudo_inverse_rule}.  

More recently, Dense Associative Memories \parencite{krotov_dense_2016}, or modern Hopfield networks \parencite{ramsauer_hopfield_2020}, have generalized the classical Hopfield network and led to a dramatic increase in memory storage capacity. This family of recurrent network architectures \parencite{krotov_large_2021} introduces a hidden layer that projects recurrently to the input layer (\autoref{fig:mhn}). Depending on the nonlinear activation function in the hidden layer, this family of networks offers a storage capacity that scales according to a power law ($\sim N^k$) or even exponentially ($\sim 2^{N/2}$) \parencite{demircigil_model_2017} with the number of input neurons (but still limited by the number of neurons in the hidden layer). However, biologically plausible plasticity rules that can implement this scaling are unknown, and instead these networks most commonly learn using gradient descent (see \autoref{sec:mlp}).

\subsection{Recognition}\label{sec:recognition}

\begin{figure}
    \begin{subfigure}{0.325\textwidth}
        \centering
        \includegraphics[height=1.5in]{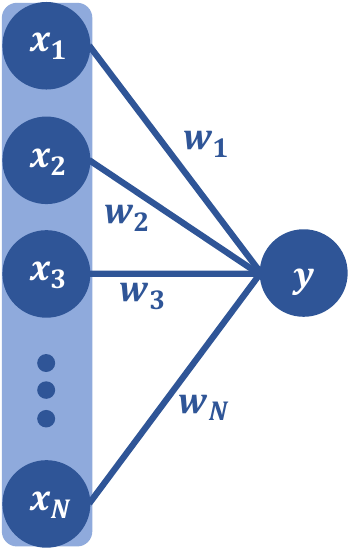}
        \caption{}    
        \label{fig:perceptron}
    \end{subfigure}
    \hfill
    \begin{subfigure}{0.325\textwidth}
        \centering
        \includegraphics[height=1.5in]{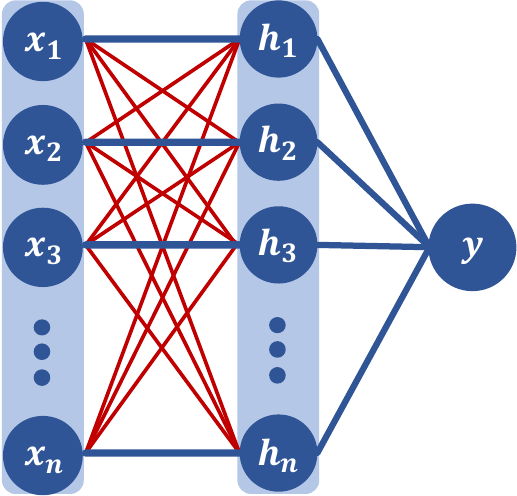}
        \caption{}
        \label{fig:bogacz}
    \end{subfigure}    
    \hfill
    \begin{subfigure}{0.325\textwidth}
        \centering
        \includegraphics[height=1.5in]{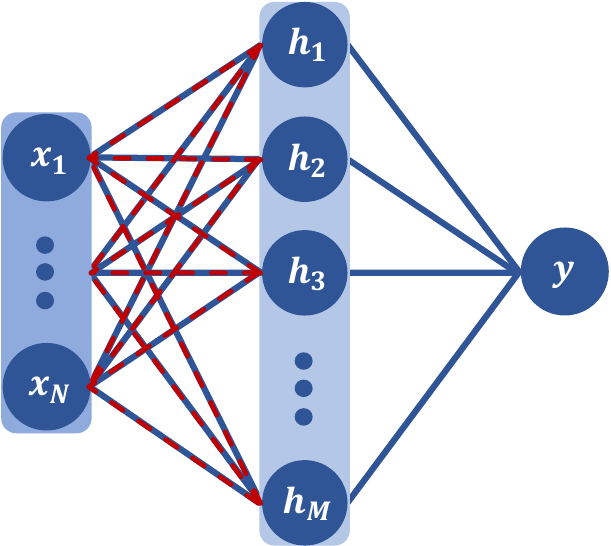}
        \caption{}
        \label{fig:hebbff}
    \end{subfigure}        
    \caption{(a) Feedforward network with a single output neuron. Depending on the plasticity rule and the input data, this architecture can be used to perform familiarity detection (\autoref{sec:recognition}), classification (\autoref{sec:supervised}), or principal components analysis (\autoref{sec:unsupervised}). (b,c) Feedforward network with one hidden layer for familiarity detection. Plastic weights are indicated in red, fixed weights in blue. Adapted from \parencite{bogacz_restricted_2002} and \parencite{tyulmankov_meta-learning_2022}, respectively.}
    \label{fig:familiarity}
\end{figure}

Often, rather than the full reconstruction of a stored item as with recall, it is sufficient to simply report whether it has been encountered in the past. Recognition memory (also referred to as familiarity or, conversely, novelty detection) accounts for common experiences such as seeing someone whose face you recognize, but don't remember where you've met them; or starting a movie and realizing you've already seen it, while still being unable to recall how it ends.%

The simplest familiarity detection network uses a feedforward architecture with a single output neuron (\autoref{fig:perceptron}). A non-Hebbian plasticity rule, controlled by the presynaptic neuron activity,
\begin{equation}\label{eq:pre_only_recog}
    \Delta w_i = \frac{1}{N}x_i
\end{equation}
ensures that the postsynaptic neuron is likely to fire for a familiar input, since a familiar pattern $\v{x}$ will be more correlated with the synaptic weight vector $\v{w}$ than a random one \parencite{bogacz_high_1999}. Recognition can also be performed by a Hopfield network \parencite{greve_optimal_2009}, although the readout involves explicitly computing its energy function, which requires a more complex network topology (\autoref{fig:bogacz}) \parencite{bogacz_model_2001}. %

An alternative, perhaps more surprising plasticity rule for recognition memory is the anti-Hebbian rule \parencite{bogacz_restricted_2002, tyulmankov_meta-learning_2022}:
\begin{equation}
        \Delta w_{ij} = -\eta x_j h_i 
\end{equation}
for $\eta>0$, where $h_i$ is the firing rate of a neuron in an intermediate "hidden" layer that receives efferents from the input layer and projects to a single output neuron (Figures \ref{fig:bogacz}, \ref{fig:hebbff}). In this case, "cells that fire together, wire \textit{apart}", meaning that the postsynaptic neurons will be suppressed for a familiar input relative to a novel one -- a biological phenomenon known as repetition suppression. This has particular benefits when inputs are correlated, as the network will tend to suppress common features and extract the components unique to the individual patterns \parencite{bogacz_comparison_2003}. In fact, for the network architecture in \autoref{fig:hebbff}, an anti-Hebbian rule reliably outperforms a Hebbian rule, and can be reliably found through optimization \parencite{tyulmankov_meta-learning_2022}. These networks might serve as models of the perirhinal cortex \parencite{bogacz_comparison_2003} or the inferotemporal cortex \parencite{tyulmankov_meta-learning_2022}. Both brain structures are implicated in recognition memory \parencite{xiang_differential_1998, meyer_single-exposure_2018}, and exhibit neural activity that is captured by these models. 

\subsubsection{Recognition memory capacity}\label{sec:recog_capacity}

Through a signal-to-noise ratio analysis \parencite[\eg,][]{hopfield_neural_1982, bogacz_high_1999},
it can be shown that the capacity of familiarity discrimination networks scales in proportion to the number of synapses, with the constant of proportionality varying depending on the plasticity rule and network architecture \parencite{bogacz_comparison_2003}. This is reasonable from an information perspective: for every synapse, there are ${\sim} 1$ bits stored. To recall an item (\autoref{sec:hopfield_capacity}), we need to store ${\sim} N$ bits of information (${\sim} 1$ bits for every entry in the vector $\xi$); to recognize it, we only need to store ${\sim} 1$ bits (whether it has been encountered previously). Therefore, while a Hopfield network with $N^2$ synapses can retrieve (recall) ${\sim} N$ items of $N$ bits each, a familiarity network with the same number of synapses can retrieve (recognize) ${\sim} N^2$ items of ${\sim} 1$ bits each. Moreover, this is consistent with the exceptional recognition memory capacity measured in humans \parencite{standing_learning_1973, brady_visual_2008}.

Unlike associative memory, however, familiarity discrimination networks are less affected by the sparsity of the input representations \parencite{bogacz_restricted_2002}. On the other hand, correlations among the inputs may play a significant role, depending on the specific network architecture and plasticity rule \parencite{bogacz_comparison_2003}.

\section{Learning paradigms}\label{sec:learning} 
In contrast to \autoref{sec:memory}, where the focus was the simple memorization of inputs, here we consider what may be more accurately described as "learning". In the context of computer science, "statistical learning", or more broadly "machine learning" refers to methods that rely on statistical techniques to understand and make predictions based on data \parencite{hastie_elements_2009}. In psychology, the related concept of "statistical learning" describes biological organisms' ability to extract regularities about their environment over time \parencite{schapiro_statistical_2015}. Learning can be broken down into three major categories -- supervised, unsupervised, and reinforcement learning, hypothesized to occur in the cerebellum, cortex, and basal ganglia, respectively \parencite{doya_what_1999}, although it is likely that these learning paradigms are used throughout the brain in various combinations.

\subsection{Supervised learning}\label{sec:supervised}
We begin with supervised learning, where the goal of the system is to estimate a function based on input-output data samples \parencite{knudsen_supervised_1994}. For example, given an input image, this can be a discrete-valued function (classification) where its output might correspond to the class of the object represented, or a continuous-valued function (regression) where its output might be the distance to a target.

\subsubsection{Perceptron}\label{sec:perceptron}
One of the foundational supervised learning algorithms for neural networks is the perceptron algorithm \parencite{rosenblatt_perceptron_1958}. The network consists of $N$ input neurons connected to a single output neuron (\autoref{fig:perceptron}). Given a dataset of input/output pairs $\{(\v{x}^1, y^1), \ldots, (\v{x}^P, y^P)\}$, the network's objective is to learn a binary classification, mapping inputs $\v{x}^p$ to one of two possible classes denoted by $y^p \in \{+1, -1\}$. For any input/output pair $(\v{x}, y)$, the synaptic weights are updated according to the error-driven plasticity rule:
\begin{equation}\label{eq:perceptron}
\Delta w_i = \eta x_i (y-\hat{y})
\end{equation}
where $\eta$ is a learning rate (which, in this case, can be any positive value) and
\begin{equation}\label{eq:perceptron_output}
\hat{y} = \sgn \left( \sum_{i=1}^N w_i x_i + b \right)
\end{equation}
is the postsynaptic firing rate corresponding to the network's estimate of the desired output $y$, given presynaptic input $\v{x}$. Note that this learning rule follows a similar structure to the Hebbian learning rule (\autoref{eq:hebb_mult}). However, rather than postsynaptic activity $\hat{y}$ alone, the synaptic update is determined by the error $(y-\hat{y})$. 
If the desired output is $y=+1$ and the network’s output is $\hat{y}=-1$, the synapses are potentiated, $\Delta w_i \propto x_i$; 
vice versa, $y=-1, \hat{y}=+1$, they are depressed, so $\Delta w_i \propto -x_i$. 
If the input pattern $\v{x}$ is classified correctly $(\hat{y}=y)$, there is no update. It is trivial to extend this learning rule to a network with multiple output units, mapping inputs $\v{x}$ to vector-valued outputs $\v{y}$ -- each postsynaptic neuron $y_j$ independently follows the learning rule given by \autoref{eq:perceptron} with the output $\hat{y}_j$ given by \autoref{eq:perceptron_output}.

The perceptron was subsequently suggested as an early model of learning in the cerebellum \parencite{albus_theory_1971}.\footnote{A similar model was proposed by \textcite{marr_theory_1969}, although he did not explicitly map it onto the perceptron.} In this view, thousands of cerebellar granule cells provide a (sparse) input through parallel fibers to Purkinje cells, each of which corresponds to a perceptron's output neuron. The output target for a single Purkinje cell then comes in through a climbing fiber input from the inferior olive, enabling local learning to occur through error-driven plasticity at the parallel fiber synapses. Overall, the cerebellum remains a likely candidate for a site of supervised learning in the brain \parencite{raymond_computational_2018}.

\subsubsection{Perceptron capacity}
As with recall (\autoref{sec:hopfield_capacity}) and recognition (\autoref{sec:recog_capacity}), we can study the perceptron storage capacity. In this case, we ask, given $P$ random $N$-dimensional points $\{\v{x}^1, \ldots, \v{x}^P\}$ how many of the $2^P$ "dichotomies" -- ways to assign binary labels $\{y^1, \ldots, y^P\}$ -- can the perceptron learn? The capacity $P_{max}$ is defined to be the maximum number of points such that at least half of the dichotomies are learnable by the perceptron (\ie, are linearly separable). This is a reasonable threshold, as the fraction of learnable dichotomies drops sharply from 1 to 0 as $P$ increases, particularly for large values of $N$. Using the counting function theorem \parencite{cover_geometrical_1965}, it is possible to show that, for random inputs, $P_{max}=2N$ \parencite[Ch. 5.7]{hertz_introduction_1991}. A more general approach \parencite{gardner_space_1988} can be used to generalize this result to correlated patterns \parencite[Ch. 10.2]{hertz_introduction_1991}, proving and quantifying the intuition that the capacity is higher for correlated patterns, although it nevertheless scales linearly with the number of input neurons $N$.

\subsubsection{Multilayer perceptron}\label{sec:mlp}
\begin{figure}
    \centering
    \includegraphics[width=0.6\linewidth]{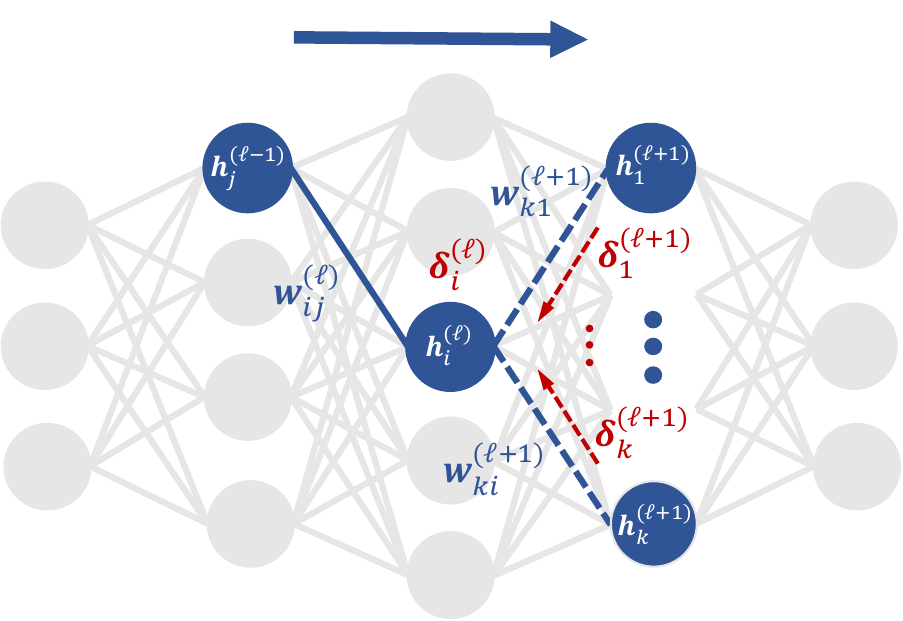}
    \caption{Multilayer perceptron, also called a feedforward neural network, or deep neural network with an input layer, three hidden layers, and an output layer. Dark blue arrow shows forward propagation for processing information. Dashed red arrows show backpropagation of errors $\delta_k^{(\ell+1)}$ to update the weight $w_{ij}^{(\ell)}$ highlighted in solid blue.}
    \label{fig:backprop}
\end{figure}

The simple one-layer perceptron, however, has a major limitation -- it is only capable of classifying patterns that are linearly separable \parencite{minsky_perceptrons_2017}. The solution to this is the inclusion of one or more "hidden" layers (\autoref{fig:backprop}). This type of architecture, known as the multilayer perceptron, feedforward neural network, or deep neural network \parencite{goodfellow2016deep} is capable of learning much more complex input-output relationships.  Even with a single (sufficiently large) hidden layer, it can approximate arbitrary functions to any desired degree of precision \parencite{cybenko_approximation_1989}.

Learning in multilayer networks, however, is much more difficult to implement. A change in a synaptic weight in an early layer affects the computation performed by its postsynaptic neuron, which in turn affects all downstream neurons. Unlike the perceptron, where multiple outputs can be learned independently, due to the nonlinear relationship between the synaptic weights and output of the neural network, it is difficult to predict a synapse's contribution to the output -- a problem known as credit assignment \parencite{richards_dendritic_2019}. For artificial neural networks, one solution is the backpropagation algorithm, proposed in this context\footnote{This algorithm was known earlier in other fields, \eg, \parencite{werbos_1974_beyond}.} by \textcite{rumelhart_learning_1986}. 

Artificial neural networks learn using a general optimization scheme known as gradient descent. We first define a metric of how well the network's output $\hat{y}(\v{x})$, a function of the input $\v{x}$, estimates the desired output $y$. This is known as a loss function, and a common choice is the quadratic (or, mean-squared error) loss: 
\begin{equation}\label{eq:mse_loss}
    \mathcal{L} = \frac{1}{2} \left( y - \hat{y}(\v{x}) \right)^2 
\end{equation}

The network parameters are then iteratively updated in such a way as to minimize the loss, which corresponds to updating them in proportion to the negative gradient of the loss function with respect to the parameters (\ie, "descending" the loss function by following the gradient). In the case of neural networks, this can be thought of as a synaptic plasticity rule of the form:
\begin{equation}\label{eq:gradient_descent}
   \Delta w_{ij}^{(\ell)} = -\eta \frac{\partial \mathcal{L}}{\partial w_{ij}^{(\ell)}} = -\eta h_j^{(\ell-1)} \delta_i^{(\ell)}
\end{equation}
for synaptic weight $w_{ij}^{(\ell)}$ in layer $\ell$ in the network (\autoref{fig:backprop}). The closed-form expression in the second equality is computed using the backpropagation algorithm, where $\delta_i^{(\ell)}$ is called the "error". At first glance, it has the familiar form of a Hebbian-like rule, where the error acts as the postsynaptic term similar to the perceptron. However, computing the error in layer $\ell$ requires knowledge of every error at layer $\ell+1$:
\begin{equation}\label{eq:backprop_error}
    \delta_i^{(\ell)} = \phi'( z_i^{(\ell)} ) \sum_k w_{ki}^{(\ell)} \delta_k^{(\ell+1)}
\end{equation}
where $z_i^{(\ell)}$ is the feedforward input to neuron $i$, and $\phi'$ is the first derivative of the nonlinearity, \ie, $h_i^{(\ell)} = \phi\left(z_i^{(\ell)}\right)$. Due to this dependence on downstream neurons (the errors are "backpropagated" through the network),  the update is no longer local to the synapse and cannot be expressed as a Hebbian plasticity rule (\autoref{eq:hebb_general}). The exception to this is the final layer of the network, where, assuming a linear output neuron and using a quadratic loss (\autoref{eq:mse_loss}), the error is given by $\delta_i^{(L)} = y_i - \hat{y}_i$ and the weight update in \autoref{eq:gradient_descent} reduces the "delta rule" \parencite{widrow1988adaptive}, which has notable similarities to the perceptron learning rule (\autoref{eq:perceptron}). Hence, the perceptron can be thought of as performing gradient descent in a biologically plausible manner. 

Backpropagation, despite its success in artificial networks \parencite{lecun_deep_2015}, has some outstanding issues as a model of learning in biological organisms \parencite{marblestone_toward_2016}. For instance, note that \autoref{eq:backprop_error} implies the existence of feedback weights that are symmetric to the feedforward weights (\autoref{fig:backprop}, dashed blue lines); such symmetry is challenging to ensure in biological neurons. Moreover, computing the error requires computing the derivative $\phi'$ of the activation function; due to the spiking nature of biological neurons, this is not well-defined. Nevertheless, biologically plausible variations have been proposed which approximate the gradient \parencite{lillicrap_backpropagation_2020, whittington_theories_2019}, although they do not scale well to more complex datasets and network architectures \parencite{bartunov_assessing_2018}. Recurrent neural networks can also learn using a similar algorithm known as backpropagation through time, but its biological plausibility has also been only partially addressed \parencite{lillicrap_backpropagation_2019}.%

\subsection{Unsupervised learning}\label{sec:unsupervised}
In many cases, a system may not be given targets onto which it must map its inputs. Its task in this case is to extract patterns from unlabeled data -- to find meaningful clusters of datapoints, to compress the dataset, or to estimate its underlying probability distribution in an unsupervised manner. Memory storage for recall or recognition (\autoref{sec:memory}) may be also considered to be a special case of unsupervised learning.%

\subsubsection{Principal component analysis}\label{sec:pca_oja}
Perhaps the most well-known unsupervised learning algorithm is principal component analysis (PCA) \parencite[see \eg,][]{greenacre_principal_2022}, the goal of which is to find a new set of coordinate axes (principal components, or PCs) that capture the most information about a dataset. For a biological organism, for example, this might be useful for extracting the most important features of its sensory input \parencite{friston_principal_1993}. 

Suppose, for simplicity, that we would like to learn only the first principal component -- the vector along the line which best fits the data, or, equivalently, the direction of highest variance. This algorithm can be implemented by a simple neural network with a Hebbian plasticity rule. In this case, we consider a linear network, where the output neuron's firing rate is a simple weighted sum of its $N$ inputs:
\begin{equation} \label{eq:oja_output}
    y = \sum_{j=1}^N w_j x_j
\end{equation}
The learning rule, known as Oja’s rule is given by \parencite{oja_simplified_1982}:
\begin{equation}\label{eq:oja}
    \Delta w_i = \eta (x_i y - y^2 w_i)
\end{equation}	
which encodes the $N$-dimensional vector corresponding to the first PC in the weights $\v{w}$. 

The Hebbian term ($x_i y$) is sufficient to learn the direction of this PC \parencite[see \eg,][]{gerstner_hebbian_2016},  although a plasticity rule with this term alone is unstable, causing the weights to grow to infinity. The second term ($-y^2 w_i$) corresponds to a synaptic decay which ensures that the norm of the weights $\bm{w}$ is equal to 1. We can equivalently rewrite the learning rule as
\begin{equation}\label{eq:oja2}
    \Delta w_i = \eta (x_i - w_i y) y
\end{equation}	
which offers a different interpretation: rather than synaptic decay, the presynaptic term receives an inhibitory current from the output neuron through synaptic feedback weights equal to the feedforward ones. This is, in a sense, a reversed version of the perceptron learning rule (\autoref{eq:perceptron}), where the error is now between the input and the back-propagated output vector \parencite[Ch. 8]{hertz_introduction_1991}.

Several extensions to Oja's rule have been proposed, with $M$ postsynaptic neurons which concurrently learn the first $M$ principal components of the data. To prevent the output neurons from all learning the same PC, the symmetry is broken through feedback connections analogous to \autoref{eq:oja2} \parencite{sanger_optimal_1989, oja_neural_1989} or lateral inhibition in the output layer, with the lateral weights learned through anti-Hebbian plasticity \parencite{foldiak_adaptive_1989, rubner_development_1990, rubner_self-organizing_1989}.

\subsection{Reinforcement learning}
Reinforcement learning (RL) can be considered to be a weaker form of supervised learning, where the model's goal is, as with supervised learning, to construct an input-output relationship, but is not given an explicit target \parencite{sutton_reinforcement_2018}. Instead, the model receives a scalar reinforcement signal known as the "reward", which tells it only whether it is correctly performing the desired task. Most commonly, rather than being given a fixed dataset, the model -- referred to as an "agent" -- interacts with an "environment" -- a discrete or continuous set of states characterized by transitions among them\footnote{This introduces an additional complication of the agent needing to explore the environment to get better data. This, however, may be at odds with the immediate goal of maximizing reward, a dilemma known as the "explore-exploit tradeoff".}. In any given state (the model's input, $x$), the agent must choose an action to perform (the output, $y$), which determines the transition to the next state, as well as the reward received, if any. This function from states to actions (or, more generally, the probability of actions) is known as the "policy", denoted\footnote{This notation denotes the conditional probability of choosing action $y$ given the state $x$. The probability density function is denoted by $\pi(\cdot|\cdot)$ as is common in reinforcement learning, rather than the usual $p(\cdot|\cdot)$.} by $\pi(y|x)$. In general, the transitions and rewards can also be stochastic, and an agent's actions determine the probability of a state transition.%

\subsubsection{Immediate reinforcement}
RL serves as a model of reward-driven learning and decision-making in animals \parencite{niv_reinforcement_2009}. As a simple example, suppose an agent is navigating an environment where its current state is given by a vector of neural firing rates $\v{x}$ -- for example, its location encoded by place cells in the hippocampus of a rat \parencite{foster_model_2000}, or its sensory input of an odor encoded by Kenyon cells in the mushroom body of a fruit fly \parencite{rajagopalan_reward_2023}. Modeling the agent as a one-layer neural network with $M$ output neurons, its corresponding action is represented by a "one-hot" vector $\v{y}$, where the $i\th$ neuron is active ($y_i=1$) if the agent takes action $i$, and silent ($y_i=0$) otherwise. Unlike the networks described in previous sections, this action is chosen stochastically, in proportion to probabilities $\pi_i$, which define the policy $\pi(\v{y}|\v{x})$, and are computed by (deterministically) propagating the input $\v{x}$ through the network.\footnote{Care must be taken to ensure that the network produces valid probabilities, \ie, $\sum_{i=1}^M \pi_i=1$. Typically this is done with a softmax function $\pi_i = e^{z_i} / \sum_{k=1}^M e^{z_k}$ with $z_k = \sum_{j=1}^N w_{kj} x_j$, which can be biologically interpreted as a "collective" nonlinearity \parencite{krotov_hierarchical_2021, snow_biological_2022}, \ie, $\v{\pi} = \softmax(\v{W}\v{x})$.} After taking the action, the agent receives a reward $r$. Learning can then occur through reward-modulated Hebbian plasticity, where the reward signal gates the synaptic update, controlling its sign, magnitude, or whether one occurs at all \parencite{fremaux_neuromodulated_2016}: 
\begin{equation}\label{eq:reward_hebbian}
    \Delta w_{ij} = r x_j y_i
\end{equation}
where $r$ represents the reward received from the environment, delivered to a biological synapse, for example, in the form of a neuromodulator like dopamine. More commonly, however, the deviation from the expected reward $(r-\bar{r})$ is used in its place. Thus, if a particular action $y_i$ in state $\v{x}$ leads to a higher-than-expected reward (\eg, $r=+1$), that state becomes more strongly associated with that action. If the action leads to a lower-than-expected reward or a punishment (\eg, $r=-1$), the association weakens. %

Using techniques from deep learning, this idea can be generalized to arbitrary neural network architectures beyond one layer, including deep feedforward or recurrent neural networks. This forms the foundation of "deep reinforcement learning", which has enabled modeling more complex learning scenarios and neural representations than classical RL permits \parencite{botvinick_deep_2020, jensen_introduction_2024}. A family of algorithms known as policy gradient methods directly optimize a policy $\pi(y|x)$ to maximize the expected reward $\mathbb{E}_\pi[r]$ that an agent receives, where the policy is defined by a deep or recurrent neural network with weights $\{w\}$. To do so, we simply perform gradient ascent, using the expected reward as the (negative) loss function. It can be shown \parencite{williams_simple_1992} that the update rule which approximates this gradient in an unbiased way is given by:
\begin{equation}\label{eq:reinforce}
    \Delta w = \eta r\, \frac{\partial \log \pi(y|x)}{\partial w}
\end{equation}
where the final derivative term is calculated using backpropagation or backpropagation-through-time (\autoref{sec:mlp}). The interpretation of this update equation as a biological plasticity rule suffers the same issues as
backpropagation, but can nevertheless lead to useful insights into biological learning \parencite{wang_prefrontal_2018, jensen2024recurrent, merel_deep_2019}.

\subsubsection{Delayed reinforcement}
More generally, a reward may not be delivered immediately after a certain action; it may depend on actions taken several time steps ago, specifically the sequence of actions that led to a rewarded state. Actions that lead to a reward in the future, however, should still be reinforced, which leads us to seek out a surrogate quantity that can be used in place of a reward to update the policy. The problem of attributing the contribution of each of this series of actions to the received reward is known as temporal credit assignment (in contrast to structural credit assignment in the context of backpropagation). 

The classical solution to this problem is known as the temporal difference (TD) algorithm \parencite{sutton_1988_learning}. As a surrogate for the reward $r$, the agent learns the "value" $v(\v{x})$ of each state $\v{x}$ -- the expected (future) reward it can accumulate starting from that state. As the agent explores the environment, for each consecutive pair of states $\v{x}^{(t)}, \v{x}^{(t+1)}$ that it visits, the value is updated in proportion to a quantity known as the TD error:
\begin{equation}\label{eq:TD_error}
    \delta^{(t)} = [r^{(t+1)} + v(\v{x}^{(t+1)})] - v(\v{x}^{(t)})
\end{equation}
and the value update is given by $\Delta v(\v{x}^{(t)}) = \eta \delta^{(t)}$. This learning rule can be understood from a simple self-consistency relationship: the value (i.e. expected future reward) of the current state $v(\v{x}^{(t)})$ must be equal to the value of the next one $v(\v{x}^{(t+1)})$, plus the actual reward $r^{(t+1)}$ received in that state.

The TD learning algorithm is particularly significant in neuroscience as it has found strong experimental support. \textcite{schultz_neural_1997} demonstrated that dopaminergic neurons in the ventral tegmental area of a monkey trained on delayed stimulus-reward pairings fire in response to the reward itself early in training, but gradually shift their response to fire at the presentation of the stimulus as training continues. Moreover, if the reward is omitted late in training, the dopamine neurons suppress their firing rate at the expected reward time, indicating a "reward prediction error". This is exactly the predicted behavior of the TD error over the course of learning, suggesting that it is encoded precisely by these neurons.

In a neural architecture \parencite{barto1983neuronlike, foster_model_2000, takahashi2008silencing}, the value function can explicitly be computed with a network known as a "critic". Using a linear one-layer network with weights $w_i^c$ as the critic, the value of state $\v{x}$ is given by 
\begin{equation}
    v(\v{x}) = \sum_i w_i^c x_i
\end{equation}
so that the value update $\Delta v(\v{x}^{(t)})$ corresponds to a Hebbian-like weight update given by:
\begin{equation}
\Delta w_i^c = \eta \delta^{(t)} x_i^{(t)}    
\end{equation}
using the TD error $\delta^{(t)}$ from \autoref{eq:TD_error}.%

The "actor" network with weights $w_{ij}$ implements the policy and subsequently selects an action as described in the previous section. Learning in the actor network, however, is done using the TD error computed by the critic instead of directly using the reward (\textit{cf.} \autoref{eq:reward_hebbian}):
\begin{equation}
    \Delta w_{ij} = \delta^{(t)} x_j^{(t)} y_i^{(t)}
\end{equation}
This ensures that (assuming the critic has correctly learned the values) actions $y_i^{(t)}$ which result in better-than-expected values in state $\v{x}^{(t)}$ are reinforced, and those that result in worse-than-expected values are suppressed. At the same time, using the values rather than immediate rewards enables temporal credit assignment, helping select actions which bring the agent closer to states where an actual reward is received. Such actor-critic architectures are often considered as models of the basal ganglia \parencite{joel_actorcritic_2002}.

The actor-critic architecture can similarly be extended to the deep RL setting. As before, the policy network serves as the actor, but we include a critic in this architecture -- another deep neural network trained with gradient ascent to estimate the value of a state. Now, rather than using the reward $r^{(t)}$ in updating the policy, we can again replace it with the TD error, this time computed using the critic's estimate of the state values and enabling temporal credit assignment (\textit{cf.} \autoref{eq:reinforce}:
\begin{equation}
    \Delta w = \eta \delta^{(t)} \frac{\partial \log \pi(y^{(t)}|x^{(t)})}{\partial w}
\end{equation}
which, heuristically, gives the modern "advantage actor-critic" algorithm \parencite{sutton_reinforcement_2018}. The interpretation of this update equation as a biological plasticity rule suffers the same issues as backpropagation, but can nevertheless lead to useful insights into learning.

\section{Synaptic complexity}

Biological synapses have a vast collection of machinery that enables adaptation beyond simple Hebbian-like updates, including mechanisms to ensure stability of the synaptic weights, control the average neural activity in a neural circuit, or enable more sophisticated learning paradigms. We now turn to models that introduce additional complexity to a synaptic plasticity rule to implement some of these features. 

\subsection{Synaptic decay}\label{sec:decay}
As discussed in \autoref{sec:pca_oja}, the simple Hebbian learning rule (\autoref{eq:hebb_mult}) is inherently unstable. To prevent synaptic weights from growing arbitrarily, additional mechanisms are required \parencite{abbott_synaptic_2000}. One mechanism for this with experimental support \parencite{turrigiano1998activity} is synaptic scaling through a multiplicative factor. An example of this is Oja's rule (\autoref{eq:oja}), where the weights decay at a rate proportional to their strength, as well as to the network's output, ensuring that the full vector of synaptic weights remains normalized. An even simpler mechanism would be an exponential decay that is independent of neuronal activity \parencite{miller_role_1994}:
\begin{equation}\label{eq:weight_decay}
    \Delta w_{ij} = f(x_j, y_i) - \lambda w_{ij}
\end{equation}
where $0 < \lambda < 1$ is the decay rate. Note that this fits within the family of Hebbian plasticity rules, provided $f$ is Hebbian as well, since $w_{ij}$ depends only on the history of $x_j$ and $y_i$. Synaptic decay enables a gradual forgetting of stored information, erasing memories that have not been recently accessed and are therefore less likely to be of importance, either for recall \parencite{mezard1986solvable} or recognition \parencite{tyulmankov_meta-learning_2022}. This helps models avoid the "blackout catastrophe" phenomenon \parencite{amit_modeling_1989}, where a network exceeding its memory capacity becomes incapable of retrieving any of its stored memories. 

Synaptic decay also plays an important role in gradient-based optimization with backpropagation (\autoref{sec:supervised}). This corresponds to adding a penalty to the loss function (\autoref{eq:mse_loss}), known as a regularization term: 
\begin{equation}\label{eq:loss_reg}
    \mathcal{L}_r = \mathcal{L} + \frac{\lambda}{2}\sum_{i} w_{i}^2
\end{equation}
where $\mathcal{L}$ is the original loss and the summation in the second term runs over all of the weights $w_{i}$ in all layers in the network. Differentiating this loss gives us the gradient descent plasticity rule (\autoref{eq:gradient_descent}), with an additional weight decay term as in \autoref{eq:weight_decay}:
\begin{equation}
   \Delta w_{i} = -\eta \frac{d\mathcal{L}}{dw_{i}} - \lambda w_{i}
\end{equation}
This regularization approach is known as ridge regression \parencite{hoerl_ridge_1970}. From the perspective of the loss function, this sum-of-squares penalizes having large-magnitude synaptic weights in the network, encouraging the network to distribute its learning across all its weights rather than strengthening only a few. In this view, the hyperparameter $\lambda$ (often chosen by trial-and-error\footnote{Formally this is known as hyperparameter optimization, including techniques such as cross-validated grid search or gradient-based methods \parencite{yang2020hyperparameter}.}) controls the relative strength of the penalty rather than the decay rate, although they are mathematically equivalent. Other regularization terms are possible, such as a sum-of-absolute-values, known as the "lasso" \parencite{tibshirani_regression_1996}, corresponding to a subtractive weight decay \parencite{miller_role_1994}, and encouraging sparsity in the parameters.%

\subsection{Learning signals}\label{sec:learning_signals}
Another source of synaptic complexity is the learning signal that is available at the synapse. In \autoref{sec:recognition}, we discuss a plasticity rule that can be used to perform familiarity detection by updating the synapse according to presynaptic activity alone (\autoref{eq:pre_only_recog}). With Hebbian plasticity, the synapse is updated according to both the pre- and postsynaptic activity, improving the storage capacity, and enabling other learning tasks. Even more sophisticated learning paradigms are possible if a third signal is present:
\begin{equation}
    \Delta w_{ij} = f(\mu, x_j, y_i)
\end{equation}
where $\mu$ is a scalar signal that is common to all synapses in the network.

One example of this is reward-modulated plasticity (\autoref{eq:reward_hebbian}), where a global reward signal gates the induction of plasticity, or controls its sign (potentiation vs. depression). Another such gating signal might be novelty, as it is likely to be important to store stimuli that have not been previously encountered. More generally, a network can learn to self-generate a modulatory signal based on its own neural activity \parencite{miconi_backpropamine_2019}. These gating signals induce learning within an entire neural microcircuit, and are therefore referred to as "global" third factors. They can be implemented in the brain by dopamine, signaling reward, or acetylcholine, signaling novelty; these and other neuromodulators such as noradrenaline or serotonin have been implicated in the induction of plasticity in the hippocampus and sensory cortices \parencite{kusmierz_learning_2017, fremaux_neuromodulated_2016}. 

Global third factors, however, lack the specificity to perform supervised learning with high-dimensional targets \parencite{gerstner_eligibility_2018}. To account for this, we can consider "local" third factors, which can be unique for every neuron or even every synapse: 
\begin{equation}
    \Delta w_{ij} = f(\gamma_{i}, x_j, y_i) \text{ or } \Delta w_{ij} = f(\gamma_{ij}, x_j, y_i)
\end{equation}
where the third factor is a synapse-specific ($\gamma_{ij}$) or neuron-specific ($\gamma_{i}$) signal. These might correspond, for example, to dendritic plateau potentials, which have been shown to induce the formation of place cells in the hippocampus \parencite{bittner_behavioral_2017}, or feedback signals from higher brain areas \parencite{roelfsema_control_2018}. Such signals have also been hypothesized to be the substrate for implementing a version of the backpropagation algorithm in the brain \parencite{guerguiev_towards_2017}. In the simpler associative memory setting, global and local third factors can be used in conjunction to separately gate the timing and the location of synaptic plasticity, respectively \parencite{tyulmankov_biological_2021}.

\subsection{Metaplasticity}
So far, we have discussed plasticity rules that are predetermined and fixed for the lifetime of a neural network. There are, however, a number of biological mechanisms that enable the plasticity rule itself to change over time \parencite{abraham_metaplasticity_2008}. Metaplasticity (\ie, the plasticity of plasticity) can serve as a homeostatic mechanism by balancing long-term potentiation and depression at a synapse, thereby preventing saturation of learning, and enabling the neuron to respond to a wider dynamic range of inputs. 

A classical example of this idea is the Bienenstock-Cooper-Munro (BCM) model \parencite{bienenstock_theory_1982}. The core component of this synaptic update rule is its non-monotonic, time-varying dependence on the postsynaptic activity: 
\begin{equation}
    \Delta w_i = \eta_w y(y-\theta)x_i
\end{equation}
where $\theta$ is a variable threshold. If the postsynaptic activity $y$ is below this threshold, the synapse will be depressed; above, it will be potentiated. Critically, the threshold itself depends on the postsynaptic neuron's activity history, for example a low-pass filtered version of it \parencite[Ch 8.2]{dayan_theoretical_2005}: 
\begin{equation}
    \Delta \theta = \eta_\theta (y^2 - \theta)
\end{equation}
This encourages stability, because potentiating a synapse will increase the postsynaptic firing rate $y$ which, in turn increases the threshold $\theta$, thus making it more likely for other synapses to be depressed. Conversely, depressing a synapse makes it more likely for other synapses to be potentiated. %

Metaplasticity can also serve as a mechanism to stabilize learning at the synaptic level, a process known as synaptic consolidation. %
One way to capture these dynamics is the cascade model \parencite{fusi_cascade_2005, benna_computational_2016}. Given a synapse with weight $w$, this model proposes an additional set of variables $u_k$ internal to the synapse that might, for example, represent concentrations of molecules in a biochemical process. Each variable $u_k$ evolves so as to equilibrate around a weighted average of its neighbors; in the simplest instantiation, each one  interacts with two neighbors $u_{k-1}$ and $u_{k+1}$:
\begin{equation}
    \Delta u_k = \eta_{k-1, k}(u_{k-1}-u_{k}) + \eta_{k, k+1}(u_{k}-u_{k+1})
\end{equation}
where $u_1 = w$ and is updated according to $\Delta u_1 = \Delta w + \eta_{1,2}(u_1-u_2)$. The input-driven update $\Delta w$ is given by a task-relevant plasticity rule, such as a Hebbian rule for familiarity detection \parencite{ji-an_face_2023} or a reward-based rule for reinforcement learning \parencite{kaplanis_continual_2018}. This update is propagated through the cascade of internal variables $u_k$ with rates determined by parameters $\eta_{k, k+1}$. As a result, depending on its history of inputs, the input-driven update $\Delta w$ will have different effects on the synaptic weight. For instance, if the synapse has undergone a long series of potentiation events, it will be more difficult to induce a synaptic depression. This resistance to overwriting enables longer memory lifetimes, while still allowing the network to learn. 

A related function of metaplasticity is to enable continual learning in neural networks \parencite{parisi_continual_2019, jedlicka_contributions_2022}. With the simple rules discussed so far, a neural network trained to perform one task (\eg, classification of cats vs. dogs) will lose this knowledge after being trained on a subsequent task (\eg, classification of cars vs. planes), a problem known as catastrophic forgetting. Techniques for continual learning attempt to mitigate this problem, not only for practical applications but also for understanding biological learning, as biological organisms do not suffer from this problem. One possible solution, known as "elastic weight consolidation" \parencite{kirkpatrick_overcoming_2017} or "synaptic intelligence" \parencite{ zenke_continual_2017}, relies on regularization, as discussed in \autoref{sec:decay}. In this case, however, the regularizer does not induce forgetting through uniform weight decay but rather prevents the overwriting of synaptic weights that are important for performing a previously learned task. 

Specifically, suppose a network has been trained on a task (or sequence of tasks), and the synaptic weights have converged to values $\widetilde{w}_i$. Training on a new task, defined by the loss $\mathcal{L}$, would then use the regularized loss function:
\begin{equation}
    \mathcal{L}_{r} = \mathcal{L} + \frac{1}{2} \sum_{i} u_i (w_i - \widetilde{w}_i)^2 
\end{equation}
which, using gradient descent, gives a synaptic update rule of the following form:
\begin{equation}
       \Delta w_{i} = -\eta \frac{d\mathcal{L}}{dw_{i}} - u_i (w_i -\widetilde{w}_i)
\end{equation}
which can be interpreted as a drift towards the original values $\widetilde{w}_i$. The internal synaptic variable $u_i$ indicates the importance of each synapse for the original task and therefore controls the rate of the drift. It can be computed online during learning of the original task \parencite{zenke_continual_2017}, 
once the task has been learned \parencite{kirkpatrick_overcoming_2017}, 
or set by explicit optimization \parencite{zucchet_contrastive_2022}.
In each case, the regularization term depends on the history of synaptic updates, so this update rule can be conceptualized as metaplasticity and implemented through additional biochemical mechanisms within a synapse. See \CHAP[\parencite{zenke_theories_2024}]{1003. Theories of synaptic memory consolidation and intelligent plasticity for continual learning} for an in-depth discussion of this and related techniques.

\section{Network architecture}
In some cases, to understand the biophysics and dynamics of a synaptic plasticity, it is instructive to study it at the level of a single synapse. However, to understand the storage capabilities of systems that rely on plasticity, it is necessary to study populations of synapses. The simplest approach uses the "ideal observer" model \parencite{fusi_memory_2021}, in which we consider a set of synapses, represented with weights $w_i$, without regard to the neurons which they connect or network architecture they are part of. A "memory" in this case is simply the set of incremental changes (potentiation or depression) $\Delta w_i$ in the set of synaptic weights. Subsequent synaptic updates occur randomly through ongoing activity, \eg, due to storage of memories, or random synaptic fluctuations \parencite{benna_computational_2016, fusi_cascade_2005, fusi_limits_2007, lahiri_memory_2013}, although in general they can also be determined by a specific learning paradigm such as supervised, unsupervised, or reinforcement learning \parencite{lindsey_selective_2024}. Critically, to establish whether a memory remains available for recall, the ideal observer assumes direct access to the synaptic weights rather than using neural circuitry to read them out, therefore this approach is an upper bound on the memory capabilities of a network.

In reality, however, synapses can only be read out from or written to through neural activity. Since multiple synapses project to a single neuron, they are no longer independent and cannot be updated arbitrarily as in the ideal observer model. Moreover, as in the case of recall or recognition with multi-layer or recurrent networks, input patterns $\xi$ cannot be mapped one-to-one to elements of the synaptic update $\Delta w_i$. The network architecture dictates the information that is available at a synapse for driving plasticity. For instance, given a presynaptic input, strong feedforward weights (Figures \ref{fig:bogacz}, \autoref{fig:hebbff}) can impose the postsynaptic activity pattern that is useful for storage with a Hebbian-like rule and subsequent downstream readout. Furthermore, extra information such as a third factor (\autoref{sec:learning_signals}) or a target signal (\autoref{sec:supervised}) must come in through auxiliary external inputs. Additional information for a synaptic update can come from within the neural circuit as well: extensions of Oja's rule to multiple outputs, for example, posit the existence of lateral connections in the output layer to allow the output neurons to learn unique principal components (\autoref{sec:unsupervised}); the backpropagation algorithm requires symmetric feedback connections to allow the error information to propagate through the network (\autoref{sec:supervised}). 

Furthermore, depending on the level of abstraction, a model's learning rule may be interpreted in different ways at the biophysical level. Even the simple Hebbian plasticity rule can effectively be implemented in several ways by including feedback projections or interneurons \parencite{sejnowski_building_1992}. Moreover, for simplicity, more abstract models may omit some biological restrictions such as Dale's law \parencite{eccles_electrical_1975}, which states that a neuron releases the same set of neurotransmitters at all its axon terminals. In models that do take this into account, a neuron can either excite or inhibit its postsynaptic neighbors, but not both \parencite[Ch. 7.2]{dayan_theoretical_2005}. Models that have neurons with both excitatory and inhibitory synapses therefore imply the existence of additional circuitry such as inhibitory interneurons which, if excited by a presynaptic neuron, inhibit their postsynaptic partners. This can lead to different plasticity rules with different neural architectures implementing the same computation. For example, in an abstract model, anti-Hebbian plasticity at excitatory synapses \parencite{bogacz_comparison_2003, tyulmankov_meta-learning_2022} may be equivalent to Hebbian plasticity at inhibitory synapses \parencite{schulz_generation_2020}. By extension, the same plasticity rule embedded in different neural architectures can lead to different learning outcomes \parencite{morris_synaptic_1990}.

\section{Non-synaptic learning}
Although the bulk of this chapter has focused on storage of information through synaptic updates, a body of evidence suggests that learning may also occur through non-synaptic mechanisms. For example, neurogenesis can be a source of major structural changes in a network, and is a growing subject of investigation \parencite{miller_functions_2019}; computational models may further elucidate its role in learning and memory \parencite{aimone_computational_2016}. Moreover, learning can also occur on timescales that outlast the lifetime of a single organism. In this case, it may be more precise to refer to the adaptive process as evolution, which may optimize an organism's neural architecture \parencite{floreano2008neuroevolution}, learning algorithm \parencite{confavreux_meta-learning_2020}, or both \parencite{tyulmankov_meta-learning_2022}. 

As a particularly interesting case, learning can occur without any structural network changes at all, purely within the recurrent neuronal dynamics of a network \parencite{wang_prefrontal_2018}. First, the authors use RL to train the synaptic weights in a recurrent neural network on a series of "two-arm bandit" tasks where the agent must choose between two actions, each of which has some unknown probability of being rewarded. After a fixed number of trials of a given task, a new one begins with a different pair of reward probabilities, and learning continues. After learning many such tasks, the synaptic weights are fixed, and a new task is presented. Interestingly, the network is still able to successfully adapt to this new task without changing its synaptic weights: learning occurs entirely in the recurrent network dynamics. The model has slowly (over the course of many different tasks with a shared structure) "meta-learned" a set of synaptic weights that is conducive to rapid trial-to-trial learning, offering an explanation for similar abilities long observed in animals \parencite{harlow1949formation}.

In the next section, we discuss a more well-studied variation of non-synaptic learning, in the form of gain and bias modulation.

\subsection{Intrinsic plasticity}
A source of ongoing modification to a neural network can come not only from updates to the synapse but also to the neuron itself, such as changes in its intrinsic excitability through changes in the number or the activation of various ion channels \parencite{sehgal_learning_2013, titley_toward_2017, lisman_memory_2018, debanne_plasticity_2019}. From a computational perspective, this might correspond to the gain $g$ and/or the bias $b$ of a model neuron: 
\begin{equation}
    y = \phi \left(g \sum_i w_i x_i + b\right)
\end{equation}

The gain controls a neuron's overall susceptibility to be excited by a presynaptic input. However, unlike individual synaptic weights $w_i$ which define the input patterns to which the postsynaptic neuron is most responsive, the gain is global to the entire neuron \parencite{ferguson_mechanisms_2020}. Although the gain does not play a role in discretized model neurons, such as those employing the step or sign function nonlinearity to indicate whether the neuron has fired an action potential\footnote{For example, the Heaviside step function $\Theta(x) = \begin{cases}0, x < 0 \\ 1, x \geq 0 \end{cases}$ remains unchanged if the input is scaled by any positive value $g$, \ie, $\Theta(x) = \Theta(gx)$}, it is significant in their continuous analogues -- the logistic function and hyperbolic tangent function -- as well as other continuous nonlinearities where the output indicates a firing rate, such as the rectified linear unit. In these cases, it sets the slope of the relationship between the input current and output firing rate. This can, for example, enable optimal information transmission in a neuron by allowing it to operate over its full dynamic range of outputs and can be learned continually in an unsupervised manner to adapt to changing input stimuli \parencite{stemmler_how_1999}. 

In a discretized neuron, the bias controls the threshold at which it becomes active. Even in the simple case of the perceptron (\autoref{eq:perceptron_output}), it is necessary to learn the bias together with the synaptic weights:
\begin{equation}
    \Delta b = \eta (y-\hat{y})
\end{equation}
This adaptively sets the threshold at which the output neuron fires based on the desired input-output relationship, and can account for datasets where the classification hyperplane must be offset from the origin. Similarly, in a multilayer perceptron or a recurrent neural network trained with gradient descent, biases are learned through backpropagation or backpropagation through time:
\begin{equation}
    \Delta b = -\eta \frac{d\mathcal{L}}{db}
\end{equation}

Although the (multilayer) perceptron learns both its weights and biases through supervised learning, an alternative is that the weights do not require direct access to the supervision signal. In \parencite{swinehart_supervised_2005}, the authors consider a network where the synapses are are updated according to an unsupervised Hebbian rule (\autoref{eq:hebb_mult}), but the gains and biases are updated according to either a gradient-based or reward-based error signal, enabling the network to successfully approximate a desired input-output function. 

In extreme cases, learning can be done without updating the synaptic weights at all:  \textcite{zucchet_contrastive_2022} show that given an optimally chosen but fixed set of synapses, a network can successfully perform rapid task adaptation ("few-shot learning") as well as reinforcement learning by modulating only the gains and biases of neurons. In the context of motor control, the synaptic weights do not even need to be fine-tuned, and the gain modulation does not need to be neuron-specific; it is sufficient to ensure the network's stability through a choice of synaptic weights, and learn the gain for each of several groups of neurons to perform a target movement \parencite{stroud_motor_2018}.

\section{Conclusion}
In this chapter we have discussed a diversity of synaptic plasticity rules and how they can be used to implement various learning and memory paradigms. Learning and memory can be loosely distinguished: memory stores information ("memories") verbatim for subsequent retrieval, for example in the form of recall or recognition; learning extracts statistical regularities ("knowledge") about the environment, with or without supervision or reinforcement. The models we have considered may also be putatively mapped onto brain areas, with models of recall often suggested to be descriptive of the hippocampus; recognition of the perirhinal or inferotemporal cortices; supervised learning occurring in the cerebellum; unsupervised learning in the prefrontal cortex; reinforcement learning in the basal ganglia. 

In practice, however, these systems are tightly interrelated: a variety of plasticity rules may interact to implement any one paradigm, and no paradigm must be constrained to just one brain area. The Hopfield network, for example, assumes storing uncorrelated memories. Visual inputs, however, are highly correlated both in space and in time, so the brain may learn to first decorrelate or compress them based on the statistics of the environment \parencite{benna_place_2021}, which could be done through unsupervised learning. Beyond preprocessing, different memory systems can also interact synergistically -- recall and recognition are hypothesized to be part of a dual-process system, where recall can aid recognition \parencite{norman_modeling_2003} and, interestingly, vice versa \parencite{savin_two_2011}.

Conversely, memory can be a driver for learning: the complementary learning systems theory, for example, suggests the existence of two learning systems \parencite{mcclelland_why_1995, mcclelland_integration_2020, kumaran_what_2016}. The first -- putatively, the hippocampus -- rapidly stores individual experiences in memory, similar to a Hopfield network. The second -- the neocortex -- slowly learns generalizable representations through error-based learning, driven by repeated interleaved replay of hippocampal memories \parencite{wittkuhn_replay_2021}. The details of individual episodes are discarded in a process known as systems consolidation \parencite{klinzing_mechanisms_2019}, thereby enabling the generalization of the acquired memories to new contexts \parencite{sun_organizing_2023}. Storage of individual items and compressed representations may also occur concurrently in the same network, driven by the same synaptic update algorithm, the trade-off performed through internally generated signals \parencite{tyulmankov_memorization_2023}. 

On a smaller scale, at the level of an individual synapse, multiple plasticity rules can also coexist. In the most basic case, a homeostatic component is necessary to complement the Hebbian component and prevent unbounded synaptic growth. This can be as simple as a synaptic decay, or a careful balance of Hebbian and non-Hebbian terms \parencite{zenke_diverse_2015}, with different components acting at different timescales \parencite{zenke_hebbian_2017}. Alternatively, fast changes in synaptic weights can work in conjunction with slower updates for rapid and reversible task adaptation \parencite{hinton_using_1987}, or serve as a transient memory bank in a recurrent \parencite{ba_using_2016} or feedforward \parencite{tyulmankov_meta-learning_2022} neural network whose slow weights define its task structure. 

Furthermore, although it is more challenging both analytically and computationally, we can consider networks that are not governed by a single uniform plasticity rule (or combination thereof), but rather evolve according to a distinct rule at each synapse. A simple version was proposed by \textcite{miconi_differentiable_2018}, who consider a recurrent neural network with short-term plasticity. The weights are given by $w_{ij}^{(t)} = w_{ij}^0 + \alpha_{ij} \widetilde{w}_{ij}^{(t)}$, with a fixed baseline $w_{ij}^0$ and a plastic component $\widetilde{w}_{ij}^{(t)}$ evolving according to the classical Hebbian rule. The plasticity coefficients $\alpha_{ij}$ allow for heterogeneous contributions of the Hebbian short-term plasticity throughout the network. In addition to this spatial heterogeneity, using a three-factor plasticity rule furthermore allows temporal heterogeneity, controlling the sign and/or magnitude of the Hebbian update analogously to a reward term \parencite{miconi_backpropamine_2019}.

Overall, computational models have significantly advanced our understanding of how synaptic plasticity underlies learning and memory. A variety of plasticity rules -- local rules that depend on pre- and/or postsynaptic neural activity, or more complicated ones involving third factors such as reward or error signals -- can implement a range of learning and memory paradigms, which, together, produce sophisticated adaptive behaviors. Drawing this link is often beyond the reach of experimental methods alone. As models become increasingly biologically accurate and better integrated with experimental data, they will continue to shed light on the complex mechanisms that drive learning and memory across multiple levels of organization.

\section*{Acknowledgements}
I am grateful for helpful comments, corrections, and clarifications from Stefano Fusi, Dmitry Krotov, and Kathleen Jacquerie. A special thanks to Kristopher Jensen for suggestions throughout the chapter, greatly improving the clarity of the reinforcement learning section in particular. 

\addcontentsline{toc}{section}{References}
\sloppy{\printbibliography} %
\end{document}